\documentclass[twocolumn,amssymb,amsmath,aps,pra,groupedaddress,nobibnotes,showpacs]{revtex4}
\usepackage{graphicx}

\begin{document}

\title{Quantum interference with beamlike type-II spontaneous parametric down-conversion}

\author{Yoon-Ho Kim} \email{kimy@ornl.gov}
\affiliation{Center for Engineering Science Advanced Research, Computer Science
\& Mathematics Division, Oak Ridge National Laboratory, Oak Ridge, Tennessee
37831}

\date{February 2003}

\begin{abstract}
We implement experimentally a method to generate photon-number$-$path and polarization entangled photon pairs using ``beamlike'' type-II spontaneous parametric down-conversion (SPDC), in which the signal-idler photon pairs are emitted as two separate circular beams with small emission angles rather than as two diverging cones.
\end{abstract}


\maketitle

Since late 1980's, spontaneous parametric down-conversion (SPDC) has served as a good source of correlated or entangled photon pairs for experimental studies of foundations of quantum physics and, more recently, for experiments in quantum information \cite{klyshko,mandel,shih,shih2,kwiat,kim,concent}. In type-I SPDC, photons appear as a concentric ring centered at the pump beam due to energy and momentum conservation conditions and the pair photons are selected with a set of small apertures which are positioned at two conjugate regions on the concentric ring \cite{mandel,shih,note}. In type-II SPDC, two such rings appear as the photon pairs are orthogonally polarized (one ring for each polarization). Depending on whether the orthogonally polarized photon pairs propagate collinearly or non-collinearly, one or two apertures are used to select correlated photon pairs, respectively \cite{shih2,kwiat,kim,concent}. In any case, only a small fraction of emitted photons can actually be collected and, in general, the bigger the collection aperture, the less the measured correlation. This is due to the fact that, with bigger opening of the collection apertures,  there are more probability of detecting un-correlated photons.    

Recently, beamlike type-II SPDC was reported in literature \cite{takeuchi,weinfurter1}. In beamlike type-II SPDC, the signal-idler photon pairs are emitted as two separate circular beams rather than as two diverging cones. Each beam has a Gaussian-like intensity distribution with a small divergence angle \cite{takeuchi}. The immediate advantage of beamlike type-II SPDC over usual type-I and type-II SPDC, in which small apertures are required to define conjugate spatial modes, is that nearly all emitted photons may be collected in principle (without compromising two well-defined spatial modes). As a result, beamlike type-II SPDC exhibits better pair detection efficiency than usual type-I and type-II SPDC \cite{takeuchi}. 

In this paper, we report a quantum interference experiment using beamlike type-II SPDC. Specifically, we report generation of the photon-number$-$path entangled state and the polarization-entangled state with beamlike type-II SPDC in one experimental setup. We also discuss a postselection-free Bell-state generation scheme using beamlike type-II SPDC.

\begin{figure}[b]
\includegraphics[width=3.2in]{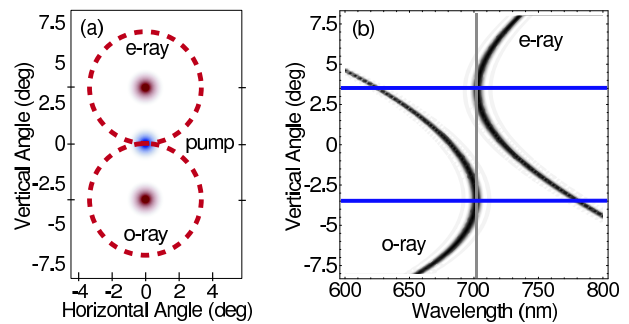}
\caption{\label{fig:tuning} (a) Two dashed lines show the emission pattern for the type-II collinear SPDC. For ``beamlike'' type-II SPDC, signal and idler photons are emitted as two separate circular beams showns as two ``blobs'' which are shown at the center of the collinear type-II SPDC emission cones. The optic axis is assumed to lie in the vertical plane and the vertical and the horizontal angles are the photon emission angles projected onto y and x axes, respectively. (b) Calculated tuning curve for beamlike type-II SPDC at 702 nm ($\theta_p\approx48.3^\circ$).  }
\end{figure}

Let us first briefly discuss how beamlike type-II SPDC can be generated. For a $\beta$-BaB$_2$O$_4$ (BBO) crystal, collinear degenerate type-II phase matching occurs when the pump beam (assumed 351.1 nm) and the optic axis of the crystal make an angle of $\theta_p=49.2^\circ$ inside the crystal \cite{shih2}. At this angle, the signal photon and the idler photon both are centered at 702.2 nm and the two emission cones or rings, shown as two dashed rings in Fig.~\ref{fig:tuning}(a), touch each other at the pump beam. The entangled photon pairs therefore propagate collinearly with the pump beam.

In the case of beamlike type-II SPDC ($\theta_p\approx48.3^\circ$), the tuning curve for both the e-ray and the o-ray are tangential to the 702.2 nm line as shown in Fig.~\ref{fig:tuning}(b). The signal and the idler photons are therefore emitted as two separate ``beams'', shown as two signal and idler ``blobs'' in Fig.~\ref{fig:tuning}(a), rather than two cones. Note that the signal-idler emission angles in beamlike type-II SPDC are fixed (in this case $\approx \pm3.5^\circ$) for a given pump and the SPDC wavaelengths. This is different from usual non-collinear type-II SPDC in which the signal-idler propagation angle may be easily adjusted by tilting the crystal in the optic axis plane.

\begin{figure}[t]
\includegraphics[width=3.2in]{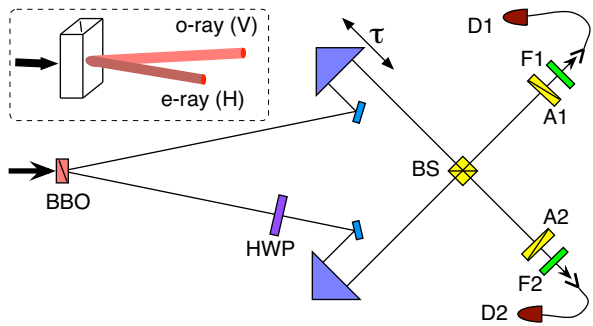}
\caption{\label{fig:setup}Outline (top view) of the experimental setup. The optic axis of the BBO crystal and the pump polarization lie in the horizontal plane.}
\end{figure}

The outline of the experimental setup can be seen in Fig.~\ref{fig:setup}. A 1 mm thick type-II BBO crystal (cut at $\theta_p=49.2^\circ$) was pumped with a 351.1 nm (horizontally polarized) argon ion laser beam. The optic axis of the crystal lied in the horizontal plane so that both the signal and the idler photons propagated parallel to the surface of the optic table. The direction of the optic axis was set as shown in the inset of Fig.~\ref{fig:setup}. The e-ray (o-ray) was therefore horizontally (vertically) polarized. A set of irises placed at $\pm3.5^\circ$ about 80 cm from the crystal helps tuning of the crystal angle for beamlike type-II SPDC generation. 

For initial alignment of the crystal angle, we placed a multi-mode fiber coupled single-photon counting detector preceeded with a 3 nm bandpass spectral filter immediately after the iris in each beam path. The distance from the crystal to the multi-mode fiber coupler was about 90 cm. Also, a horizontally (vertically) oriented polarizer was inserted in the path of e-ray (o-ray). The BBO crystal was then sligntly tilted in the horizontal direction while observing the single as well as the coincidence counts of the detectors. Only a slight detuning of the crystal angle was necessary to observe beamlike type-II SPDC. 

Similarly to the results reported in Ref.~\cite{takeuchi}, the ratio of the coincidence-count rate to the single-count rate was approximately $11.5\%\sim12\%$ when the irises were about $5\sim7$ mm in diameter. When the pump beam was slightly focused with a 1 m focal lens (located about 1 m from the detectors), the ratio rose up to $\sim15\%$ for similar size irises \cite{monken}. Experimentally, this ratio is mostly limited by the detector efficiency (EG\&G SPCM, typically 70\% at this wavelength), the fiber coupler efficiency (about 65\%, measured with a He-Ne laser), the filter transmission (approximately 55\% peak transmission), and other small optical losses. The filter transmission, therefore, is the biggest factor for the coincidence/signal ratio and it may be improved by using better designed filters and by using a thicker BBO crystal which reduces the spectral bandwidth of SPDC. Note that, if the mode-matching technique described in Ref.~\cite{weinfurter2} is used with the beamlike type-II SPDC, a much improved coincidence/single ratio (in turn better pair detection efficiency) may be possible \cite{note12}.

\begin{figure}[t]
\includegraphics[width=3.2in]{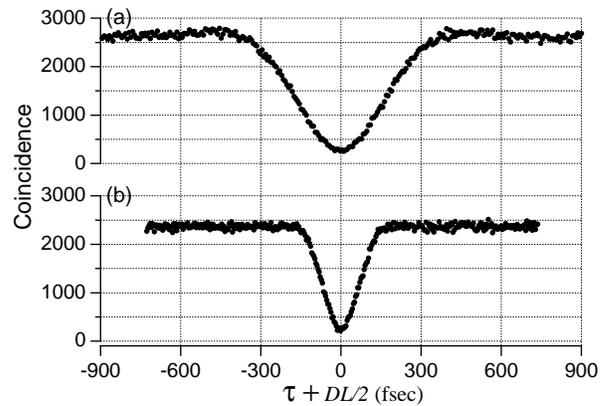}
\caption{\label{fig:dip}Experimental data with HWP angle set at $45^\circ$. (a) with 3 nm filters (5 sec), (b) with 20 nm filters (4 sec).}
\end{figure}

The interferometer for beamlike type-II SPDC consists of a half-wave plate and a 50/50 beamsplitter as shown in Fig.~\ref{fig:setup}. The horizontally polarized signal beam passed through a half-wave plate (HWP) and the signal beam and the idler beam were overlapped at the beamsplitter (BS). The delay $\tau$ between the two arms was adjusted with a computer controlled trombone prism. At each output ports of the beamsplitter, a polarization analyzer (A1 or A2), a spectral filter (F1 or F2), an iris, and a fiber-coupled single-photon detector were placed. The distance from the crystal to the detector iris was about 190 cm. The coincidence window used in this experiment was approximately 7.3 nsec.

Let us first discuss the two-photon anti-correlation dip experiment with beamlike type-II SPDC. For this measurement, the HWP was set at $45^\circ$ so that both the signal and the idler photons have the same polarization and the polarization analyzers A1 and A2 were removed from the setup. Since both photons have the same polarization, the two-photon dip effect similar to that of type-I SPDC is expected \cite{mandel}. There are, however, some subtle differences because beamlike type-II SPDC should show all the characteristics of usual type-II SPDC as far as interference effects are concerned. Therefore, the dip should be triangular in the absence of any spectral filtering and it should occur at $\tau=-DL/2$ \cite{shih2}. Here $D=1/u_i-1/u_s$, where $u_s$ ($u_i$) is the group velocity of the signal (idler) photon in the crystal and $L$ is the thickness of the crystal. In other words, complete anti-correlation dip or two-photon bunching occurs when the effective time delay through the upper arm of the interferometer is $DL/2$ shorter than that of the lower arm. 

The experimental data for the anti-correlation dip experiment are shown in Fig.~\ref{fig:dip}.  Fig.~\ref{fig:dip}(a) shows the experimental data when the spectral filters F1 and F2 had 3 nm FWHM bandwidth. The Gaussian-like shape of the two-photon dip is due to spectral filtering of SPDC by the 3 nm filters. The familiar triangular two-photon dip for type-II SPDC is evident in Fig.~\ref{fig:dip}(b) when 20 nm FWHM spectral filters were used instead. The triangular shape two-photon dip indicates that the spectrum of type-II SPDC is not affected by spectral filters and apertures used in the experiment \cite{shih2}. The two data sets have comparable coincidence count rates due to different sizes of detector irises (3 mm in diameter for Fig.~\ref{fig:dip}(a) and 2 mm for Fig.\ref{fig:dip}(b)), different measurement times (5 sec for Fig.~\ref{fig:dip}(a) and 4 sec for Fig.\ref{fig:dip}(b)), and slightly lower peak transmission of the 20 nm filters than that of 3 nm filters. 

The quantum state of the output of the beamsplitter, at the dip, can be written as $|\psi\rangle = (|2,0\rangle + |0,2\rangle)/\sqrt{2}$, which is the well-known photon-number$-$path entangled state. So far, such postselection-free photon-number$-$path entangled state was reported using type-I SPDC only in which the spectral bandwidth is very broad. The reduced bandwidth and improved pair detection efficiency of beamlike type-II SPDC maybe useful for certain quantum applications of the photon-number$-$path entangled states.

\begin{figure}[t]
\includegraphics[width=3.3in]{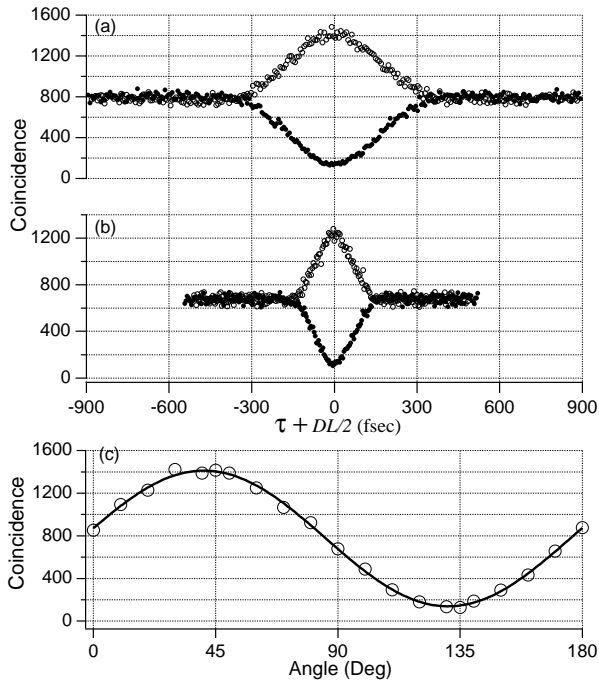}
\caption{\label{fig:peakdip}Experimental data with HWP angle set at $0^\circ$. (a) with 3 nm filters (5 sec), (b) with 20 nm filters (4 sec). Solid circles are for $\textrm{A1/A2}=45^\circ/45^\circ$ and empty circles are for $\textrm{A1/A2}=45^\circ/-45^\circ$. (c) polarization correlation measurement with 3 nm filters (5 sec).}
\end{figure}

The generation of the polarization-entangled state may be accomplished by simply setting the HWP angle at $0^\circ$, i.e., it does not do anything to the polarization state of the signal photon. In this case, the situation is very similar to that of the first Bell-inequality experiment using type-I SPDC \cite{mandel,shih}. To observe the usual coincidence dip pattern \cite{shih2}, the polarization analyzers A1 and A2 were both set at $45^\circ$ and the delay between the the two interferometer arms was scanned. This procedure was repeated with two different sets of spectral filters (with 3 nm and with 20 nm). To observe the coincidence peak, the orthogonal analyzer angles $\textrm{A1/A2} = 45^\circ/-45^\circ$ were used. 

The experimental data for this measurement can be seen in Fig.~\ref{fig:peakdip}(a) and (b). The data clearly confirm the polarization entangled state $|\psi\rangle = (|H\rangle_1|V\rangle_2 - |V\rangle_1|H\rangle_2)/\sqrt{2}$ at $\tau=-DL/2$. The demonstration of polarization correlation measurement  ($\textrm{A1}=-45^\circ$ and A2 rotated) at $\tau=-DL/2$ can be seen in Fig.~\ref{fig:peakdip}(c). Note that if $\tau\neq-DL/2$, partially-entangled partially-mixed state may be generated.

\begin{figure}[t]
\includegraphics[width=2.5in]{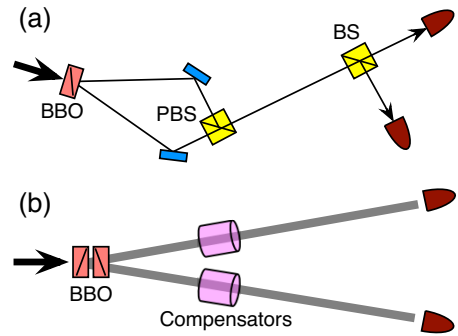}
\caption{\label{fig:proposal}Other methods to generate polarization entangled states. (a) This setup is similar to type-II collinear SPDC. (b) Postselection-free polarization entanglement scheme. }
\end{figure}

Beamlike type-II SPDC may also be used prepare the two-photon state somewhat similar to that of collinear type-II SPDC. An example can be seen in Fig.~\ref{fig:proposal}(a). Here, a polarizing beamsplitter is used instead of a beamsplitter. The resulting state after PBS, therefore, is similar to the state prepared by collinear type-II SPDC \cite{shih2} or two collinear type-I SPDC's \cite{kim,kim2}. This type of a single-mode two-photon state can be described as a three-state quantum system (qutrit) \cite{moscow} and the possibility of quantum information processing using such ternary quantum states are currently being investigated by many researchers.

In Fig.~\ref{fig:proposal}, we show a scheme to generate postselection-free polarization entangled states using beamlike type-II SPDC \cite{note2}. The first and the second crystal generates beamlike type-II SPDC photon pairs in the polarization state $|H\rangle_1|V\rangle_2$ and $|H\rangle_2|V\rangle_1$, respectively. After the photons pass through suitable compensators similar to the one used in Ref.~\cite{bitton}, the two amplitudes may become distinguishable only by their polarization properties. All four two-photon polarization Bell-states therefore may be prepared in this way. This setup, therefore, may be useful for loophole-free Bell-inequality tests if combined with other collection efficiency enhancing techniques \cite{monken,weinfurter2}. 

Note that, unlike the schemes using two orthogonally oriented type-I crystals \cite{kwiat,kim,kim2}, non-maximally entangled states cannot be directly generated using beamlike type-II SPDC. In the two-crystal scheme shown in Fig.~\ref{fig:proposal}(b), the pump beam has to be horizontally polarized and the amplitudes $|H\rangle_1|V\rangle_2$ and $|H\rangle_2|V\rangle_1$ are generated in each crystal due to the different optical axes orientations, both of which lie in the horizontal plane. Therefore, the weighting factors for each amplitudes cannot be easily varied by simply changing the polarization orientation of the pump beam. Partial reflection mirrors inserted in the beam paths, however, may allow indirect generation of non-maximally entangled states.

For ultrafast pumping, beamlike type-II SPDC suffers the same visibility loss as with usual ultrafast type-II SPDC schemes. In ultrafast type-II SPDC, the signal and the idler photons have different spectra and the pump pulse acts as a clock which in principle could be used to tell where the photon pair was created inside the crysal \cite{pulsedspdctheory}. These extra distinguishing information are the source of poor visibility in usual type-II SPDC with ultrafast pump \cite{pulsedspdcexp}. Since the distinguishing information still exists in the two-photon state of beamlike type-II SPDC, the same poor visibility is expected as well. It may, however, be possible to use the entanglement concentration (or extraction) scheme demonstrated in Ref.~\cite{concent} with the two-crystal scheme shown in Fig.~\ref{fig:proposal}(b). If the two modes in the two crystal scheme shown in Fig.~\ref{fig:proposal}(b) are made to overlap at an additional polarizing beamsplitter, the spectral or temporal distinguishing information may be effectively ``decoupled" from the polarization information of the photon pair. In this case, regardless of the pump bandwidth, a high-quality post-selection free Bell-state may be generated using beamlike type-II SPDC. 

In summary, we reported a quantum interference experiment using beamlike type-II SPDC, in which the signal and the idler photons are emitted as two separate circular beams with small emission angles rather than as two diverging cones. We also discussed methods to prepare postselection-free Bell-states using beamlike type-II SPDC. Since beamlike type-II SPDC exhibits better collection and pair detection efficiencies due to its emission property, the entangled state generated using beamlike type-II SPDC is well-suited for loophole-free Bell-inequality experiments (if combined with other detection efficiency enhancing techniques) and for many quantum information experiments involving entangled photon pairs. 

This research was supported in part by the U.S. Department of Energy, Office of Basic Energy Sciences, NSA, and the LDRD program of the Oak Ridge National Laboratory, managed for the U.S. DOE by UT-Battelle, LLC, under contract No.~DE-AC05-00OR22725.


\begin{thebibliography}{}
\bibitem{note} In collinear type-I SPDC, the SPDC ring overlaps with the pump beam. In this case, however, photon pair cannot be spatially separated unless they have different wavelengths.
\bibitem{note12} In Ref.~\cite{weinfurter2}, the authors report the coincidence/single ratio of $28.6\%$ without using interference filters. With beamlike type-II SPDC, some sort of spectral filtering is necessary because dual wavelengths exist at $\approx\pm3.5^\circ$ emission angles.
\bibitem{note2} Polarization entangled state generation scheme shown in Fig.~\ref{fig:setup}, in fact, generates post-selected polarization entangled states: i.e., the amplitudes (both photons leaving the beamsplitter via the same output port) that do not contribute to a coincidence count are ignored.
\end{thebibliography}
\end{document}